\documentclass{svjour3}                     % onecolumn (standard format)
\smartqed  % flush right qed marks, e.g. at end of proof
\usepackage[normalem]{ulem}%temporal
\usepackage{graphicx}
\usepackage{color}
\usepackage{amssymb}
\usepackage{amsmath}
 \journalname{Scientometrics}

\begin{document}

\title{Absolute and specific measures of research group excellence
%Should citation-based scientometrics be extensive or intensive?
%Intensive or extensive citation-based proxies for research evaluation
%What kind of metric do we need to measure science?
}
%\subtitle{Can we do this?}%\\ If so, write it here}

%\titlerunning{}        % if too long for running head

\author{O.~Mryglod \and R.~Kenna \and Yu.~Holovatch \and B.~Berche}

\institute{O. Mryglod \at
              Institute for Condensed Matter Physics of the NAS of Ukraine, \\
              1 Svientsitskii Str., 79011 Lviv, Ukraine\\
              \email{olesya@icmp.lviv.ua}
           \and
            R. Kenna \at
              Applied Mathematics Research Centre, Coventry University,
              Coventry, CV1 5FB, England
           \and
            Yu. Holovatch \at
              Institute for Condensed Matter Physics of the NAS of Ukraine, \\
              1 Svientsitskii Str., 79011 Lviv, Ukraine
           \and
            B. Berche \at
              Statistical Physics Group, Institut Jean Lamour,
              CNRS -- Nancy Universit\'{e} -- UPVM, \\B.P. 70239,
              F-54506 Vandoeuvre l\'{e}s Nancy Cedex, France
}

\date{Received: date / Accepted: date}

\maketitle

\begin{abstract}
A desirable goal of scientific management is to introduce, if it exists, a simple and reliable way to measure the scientific
excellence of publicly-funded research institutions and universities to serve as a basis for their ranking and financing.
While citation-based indicators and metrics are easily accessible, they are far from being universally accepted as way to automate or inform evaluation processes or to replace evaluations based on peer review.
Here we consider absolute  measurements of research excellence at an amalgamated, institutional level and specific measures of research excellence as performance per head.
Using biology research institutions in the UK as a test case, we examine the correlations between peer-review-based and citation-based  measures of research excellence on these two scales.
We find that citation-based indicators are very highly correlated
with peer-evaluated measures of group strength but are poorly correlated with group quality.
Thus, and almost paradoxically, our analysis indicates that citation counts could possibly form  a basis for deciding on how to fund research institutions but they should not be used as a basis for ranking them in terms of quality.

\keywords{scientometrics \and scientific evaluation \and higher education}
\end{abstract}

%%%%%%%%%%%%%%%%%%%%%%%%%%%%%%%%%%%%%%%%%%%%%%%%%%%%%%%%%%%%%%%%%%%%%%%%%%%%%%%%%%%%%%%%%%%%%%%%%%%%%%%%%%%%%%%%%%%%%%%%%%%%5
\section*{Introduction}
%%%%%%%%%%%%%%%%%%%%%%%%%%%%%%%%%%%%%%%%%%%%%%%%%%%%%%%%%%%%%%%%%%%%%%%%%%%%%%%%%%%%%%%%%%%%%%%%%%%%%%%%%%%%%%%%%%%%%%%%%%%%5
The problem of measuring the scientific excellence  of research institutes and universities is one of current and continued importance \cite{Bellis2009}. %
It is important to be able to detect and support the most promising research groups and to have sound and robust bases to wisely plan and invest for the future.
Metrics and indicators of scientific activity may be based on the volume of researchers, the number of published papers, the amount of citations to these publications,
the number of PhD students and on finances generated by application of scientific ideas, amongst other factors.

It is often argued that automated scientometric or bibliometric indicators may be an inadequate substitute or proxy for peer-review based evaluations of the merit of research institutes  \cite{nature}.  There is a lot of arguments  against the use of indicators in isolation, suggesting they reflect only one aspect of the research and should only be used as an adjunct to peer review (for example, see~\cite{Evidence_2010,Bellis2009}). But in spite of the problems with indicators and metrics, reasonable and cheap approaches to the evaluation of scientific productivity and quality are desirable for practical purposes, not least because institutional peer review is an expensive and time-consuming exercise \cite{Bellis2009}.

%Scientometrics evolved from the second part of the 20th century to study all aspects of scientific endeavour as a complex sociological system  \cite{Price}.
%Modern computer and information technologies facilitate the collection and analysis of vast amounts of scientometric data, such as citations, number of downloads of papers or number of coauthors.
%While it is comparably easy to use these data to automate some processes in scientific management, we still don't have any universally accepted criteria of scientific efficiency.

Here we investigate ``absolute'' and ``specific'' measurements of scientific excellence on the scale of university research groups. E.g., the {\emph{absolute}} citation count for a department is the total number of citations, irrespective of how many researchers that department contains. The corresponding {\emph{specific}} citation   count is then the average number of citations per head  (see, for example, \cite{Vinkler2001,Vinkler2003}).
In physics parlance these correspond respectively to the notions of ``extensive'' and ``intensive'' quantities and are usually represented mathematically by upper-case and lower-case symbols. That convention is borrowed here for absolute and specific measurements.
In this paper we compare these specific and absolute notions of quality and
strength with  indicators of citation impact, to
determine if the latter may be used as a reliable proxy for the
former. We consider the overall {\emph{strength}} of a research collective (a university department, research centre, group or some variant, component or amalgam of these) as measured through a peer-review evaluation scheme to be an absolute quantity and denote it by ${\mathcal{S}}$.
A peer-review measure of the {\emph{quality}}\footnote{Here and further we use terms ``quality'' and ``strength'' following the notation in \cite{Ralph2010,Ralph2011}} $s$ of a group of  researchers is then the strength per head.
Similarly, the total citation impact  ${\mathcal{I}}$  is considered here as the absolute impact of research group while its average value $i$ (calculated per head) corresponds to specific impact.
Thus, ``absolute'' refers to total institutional measurements while ``specific'' means average properties per individual. We then define {\emph{strength}} as ``volume of quality'' and {\emph{absolute impact}} as ``volume of average impact''.

In this paper, to make a quantitative comparison we use data from the British system of evaluation of research and education (\emph{Research Assessment Exercise}) and from \emph{Evidence}, a company within Thomson Reuters, one of the world's leading providers of scientific information. Using biology research groups and departments in the UK as a test case,
we show that the
citation-based  measure $i$  is not a good proxy for the peer-review measure $s$, in that these two specific measures are rather poorly correlated. However, when scaled up to the actual size of the department $N$, the {\emph{absolute}} citation impact ${\mathcal{I}}=iN$ is very strongly correlated with the
overall strength ${\mathcal{S}}=sN$ as measured by peer review. (Here and below $N$ means the number of researchers in group.)
This means that citations, if used in an informed manner, could possibly be used as a proxy for departmental or group {\emph{strength}}
particularly for large groups,
but should not be as an  estimate of research {\emph{quality}}.
When applied to large research
groups or departments in particular, this may offer an alternative or at least complement to peer review.

%%%%%%%%%%%%%%%%%%%%%%%%%%%%%%%%%%%%%%%%%%%%%%%%%%%%%%%%%%%%%%%%%%%%%%%%%%%%%%%%%%%%%%%%%%%%%%%%%%%%%%%%%%%%%%%%%%%%%%%%%%%%5
\section{Background: The research assessment exercise and the normalised citation impact}
%%%%%%%%%%%%%%%%%%%%%%%%%%%%%%%%%%%%%%%%%%%%%%%%%%%%%%%%%%%%%%%%%%%%%%%%%%%%%%%%%%%%%%%%%%%%%%%%%%%%%%%%%%%%%%%%%%%%%%%%%%%%5

For the last quarter century, the \emph{Research Assessment Exercise} (RAE) provided the premier yardstick for the measurement of scientific performance of research institutions in United Kingdom.
This introduced  an explicit and formalized process to assess the quality of research \cite{RAE_web}.
Six evaluation exercise have  occurred since the first one in 1986 and evaluation frameworks of this type  now constitute the single most important event in the UK research calender each five or so years.
For the purpose of the 2008 exercise,  each academic discipline was assigned to one of 67 units of assessment (UOA).
In order to receive quality related funding any Higher Education Institution (HEI) could make submissions to RAE in any UOA.
Using published criteria, RAE experts assess these submissions and generate graded profile for each of them.
These profiles quantify the proportion of a department's or research centre's work which falls into each of five quality bands as follows~\cite{RAE_web}:
\begin{itemize}
 \item 4*: Quality that is world-leading in terms of originality, significance and rigour;
 \item 3*: Quality that is internationally excellent in terms of originality, significance and rigour but which  falls short of the highest
standards of excellence;
 \item 2*: Quality that is recognised internationally in terms of originality, significance and rigour;
 \item 1*: Quality that is recognised nationally in terms of originality, significance and rigour;
 \item Unclassified  --  Quality that falls below the standard of nationally recognised work or work which does not meet the published definition of research for
the purposes of the assessment.
\end{itemize}
An example of a quality profile is given in Table~\ref{tab1}.
\begin{table}[!b]
\caption{An example of quality profile of RAE 2008 for the biology department of Open University$^{\ddag}$.}
\begin{center}
\small{
\begin{tabular}{|l|l|l|l|l|l|l|}
\hline
 Profile type &
\multicolumn{5}{|c|}{\parbox{40mm}{\begin{center}Quality profile \\ (percentage of research activity at each quality level)\end{center}}}\\
\hline
& 4* & 3*&  2*& 1*& unclassified
\\\hline\hline
Output & 5.3 &17.7& 46.1& 23.7& 7.2
\\\hline
Environment & 0.0 &14.7 & 50.9 & 33.2& 1.2
\\\hline
Esteem &7.1 &42.5& 41.5&3.6& 5.3\\
\hline
Overall &5.0 &20.0& 45.0&25.0& 5.0\\
\hline
\end{tabular}
} \label{tab1}
\end{center}
$^{\ddag}$An overall quality profile is constructed by summing the three separate weighted components (``Output'', ``Environment'' and ``Esteem'')
using a special cumulative rounding methodology which avoids unfair consequences that simple rounding can produce  \cite{RAE_statement_UOA14}.
\end{table}

Besides obvious benefits in terms of  prestige, marketing and publicity potential, the research quality profiles determined how much funding each university department receives in the years following the exercise.
Funding is determined by a formula combining the weighted quality scores of individual  research group or centres.
The formula is subject to regional and temporal variation, but the one introduced by the Higher Education
Funding Council for England (HEFCE) immediately subsequent to RAE 2008 was \cite{HEFCE_web}
\begin{equation}
s=p_{4*}+\frac{3}{7}p_{3*}+\frac{1}{7}p_{2*}.
\label{eq1_funding_formula}
\end{equation}
Here $p_{n*}$ represents the percentage of a team's research which was rated $n*$ and $s$ may be considered as a single {\emph{specific}} measure of a university's overall research quality in a particular discipline.
The total amount of quality related funding distributed by HEFCE to a given university post RAE is proportional to the overall  {\emph{strength}} of the submission \begin{equation}
 {\mathcal{S}} = sN,
\end{equation}
where $N$ is the number of researchers submitted from a given university.
Thus, strength  ${\mathcal{S}}$ is an {\emph{absolute}} measure of performance.

Notwithstanding various managerial tactics and manoeuvres, the RAE, and similar
exercises in other countries, is considered to be reasonably reliable since it is based
on peer-review evaluation. Despite its many drawbacks and limitations, peer review is considered by the academic community at large to be the most reliable and trustworthy scheme to evaluate the worth of curiosity-driven research, in particular.
On the other hand, peer-review-based evaluation is  expensive and time-consuming process.
In addition, the very act of measuring the scientific system distorts the very process it purports to measure.
This is Goodhart's Law, a type of socioeconomic counterpart  of the Heisenberg uncertainty principle in physics.
Therefore, it is reasonable to propose some alternatives to the RAE.
Evaluations based entirely upon citation counts are some of many possible candidate schemes (see, for example, \cite{Oppenheim1996,Oppenheim2003}).
Some authors advocate substitution of the RAE by citation counting due to claimed good correlation between the resultant rankings obtained \cite{Oppenheim1996,Holmes2001}.
But there are also authors (e.g., \cite{MacRoberts1989}) who point to different weaknesses of citation analysis such as self-citing, bias,  technical errors connected with citation data retrieval or different typical rates of citations  for various disciplines.
Although methods have been, and are being, developed to try to deal with these phenomena, no universally satisfactory  automated system has yet emerged.

So, the question arises as to how good citation-based indicators can be as proxies for peer review.
Since the last UK exercise generated quality measurements in far more detail than previous RAEs, and since subsequent evaluation frameworks in the UK will amalgamate some of the UOA's used in 2008 into fewer, larger subject groupings, RAE 2008 provides a window of great detail in which answers to this question may be sought.
Having precise numbers assigned to each institution for the 67 disciplines it is possible to perform accurate comparisons between RAE peer-review evaluation and citation scores. Thus we arrive at the main questions we wish to address herein: how do RAE~2008 scores correlate with citation rates?

As already noticed before, to get the citation-based measure, we use the data provided by \emph{Evidence}. This company offers a service analysing research performance  tailored to individual client requirements \cite{Evidence_web}. We consider the so-called {\emph{normalised citation impact}} (NCI) $i$  used by {\emph{Evidence}} as a coefficient of departmental performance in a given
discipline. %

{\emph{Evidence}}  calculate the
 NCI using data from Thomson Reuters databases \cite{Evidence_2010,Evidence_2011}.
Similarly to Relative Citation Rate (RCR) (i.e., \cite{Schubert1996}), NCI is a relative measure as it is calculated by comparing to a mean or expected citation rate. On the other hand, this is a specific measure of citation impact because it is averaged to be a measure of impact for research group.
It has long been known that citation rates are different for different disciplines. The main advantage of \emph{Evidence} calculations
is non-trivial normalisation of citation counts between different academic disciplines.
To achieve  this, the total citation count for each paper can
be normalised to an average citations per paper for the year of
publication and either the field or journal in which the paper was
published (the so-called ``rebasing'' of the citation count). The
normalised value is known as the NCI
\cite{Evidence_2011}. Only the four papers per individual which were submitted to RAE
2008 were taken into account by {\emph{Evidence}} in order to calculate the average NCI for research groups \cite{Evidence_2011}. The RAE does not emplot a comparable
normalisation process and this is a serious and acknowledged
weakness of that peer evaluation exercise \cite{KBnormalization}.

Thus, the NCI may be considered as a {\emph{specific}} measure of
the research impact of a department in a given field and we denote
it by $i$. The corresponding {\emph{absolute}} measure of impact
(the total impact of the department or group) is denoted by
${\mathcal{I}}$. The relationship between the two is
\begin{equation}
{\mathcal{I}}=iN.
\end{equation}
In the following, we compare the {\emph{specific}} indicators of quality and impact $s$ and $i$ as
measures of the average strength and impact of the group or department {\emph{per individual}} contained within it. We also compare the counterpart {\emph{absolute}} measurements ${\mathcal{S}}$ and ${\mathcal{I}}$ as measures of the overall strength and total impact of the group as a whole.
We use the biology research sector in the UK as a test case.
We show that while the specific measures $s$ and $i$ are only weakly correlated, the absolute ones ${\mathcal{S}}$ and ${\mathcal{I}}$ are strongly aligned.
The analysis thus suggests that citations form a poor basis for measuring institutional research quality but could form a good basis for measuring strength.
Since quality related funding as defined by HEFCE is proportional to departmental strength rather than quality, the citation-based measures may offer a reasonable and
cost effective way to decide on funding while ameliorating the negative effects of Goodhart's law.

We have to note that although RAE assessments as well as the Evidence results are used for practical purposes, they are still subject to debate within the scientometric community. This is an additional reason to compare them.

%%%%%%%%%%%%%%%%%%%%%%%%%%%%%%%%%%%%%%%%%%%%%%%%%%%%%%%%%%%%%%%%%%%%%%%%%%%%%%%%%%%%%%%%%%%%%%%%%%%%%%%%%%%%%%%%%%%%%%%%%%%%5
\section{Weak correlation between specific measures of quality and impact}
%%%%%%%%%%%%%%%%%%%%%%%%%%%%%%%%%%%%%%%%%%%%%%%%%%%%%%%%%%%%%%%%%%%%%%%%%%%%%%%%%%%%%%%%%%%%%%%%%%%%%%%%%%%%%%%%%%%%%%%%%%%%5

%
\begin{table}[!b]
\caption{The ranking  of UK  biology departments using the RAE
2008 scores $s$ and the NCI-based scores~$i$.}
\begin{center}
\begin{tabular}{|l|l|l|l|l|}
\hline
HEI & \parbox[t]{15mm}{Average quality, $s$}& \parbox[t]{10mm}{Ranking by $s$}& \parbox[t]{15mm}{Average NCI, $i$}&\parbox[t]{10mm}{Ranking by $i$}\\
\hline \hline
Institute of Cancer Research&61.43&1&2.26&1
\\ University of Manchester&46.43&2&1.32&14
\\ University of Dundee&45.71&3&2.06&2
\\ University of Sheffield&45.00&4&1.25&17
\\ University of York&44.29&5&1.67&5
\\ Imperial College London&42.86&6.5&1.56&7
\\ King's College London&42.86&6.5&1.45&10
\\ Royal Holloway. University of London&42.14&8&1.05&31
\\ University of Cambridge&41.43&9&1.85&3
\\ University of Leeds&39.29&10&1.32&15
\\ University of Edinburgh&38.57&11.5&1.53&8
\\ University of Newcastle upon Tyne&38.57&11.5&1.20&22
\\ University of Glasgow&37.14&14&1.23&19
\\ Cardiff University&37.14&14&1.18&24
\\ University of Aberdeen&37.14&14&1.02&34
\\ University of St Andrews&33.57&16&1.13&26
\\ University of Bath&32.86&19.5&1.34&13
\\ University of Durham&32.86&19.5&1.27&16
\\ University of Birmingham&32.86&19.5&1.25&18
\\ University of Nottingham&32.86&19.5&1.13&27
\\ University of East Anglia&32.86&19.5&1.09&29
\\ University of Exeter&32.86&19.5&0.84&38
\\ University of Southampton&32.14&23.5&1.15&25
\\ University of Warwick&32.14&23.5&1.09&28
\\ University of Leicester&30.71&25&1.63&6
\\ University of Liverpool&29.29&26&1.04&32
\\ University of Essex&26.43&27.5&1.46&9
\\ Queen Mary. University of London&26.43&27.5&1.19&23
\\ University of Sussex&24.29&29.5&1.75&4
\\ University of Reading&24.29&29.5&1.04&33
\\ University of Kent&23.57&31&0.88&35
\\ Queen's University Belfast&22.86&32&0.82&39
\\ Bangor University&21.43&33&0.84&37
\\ Open University&20.00&34&0.88&36
\\ Oxford Brookes University&17.86&35.5&1.22&21
\\ University of Plymouth&17.86&35.5&0.64&42
\\ University of Hull&17.14&37&1.43&11
\\ Cranfield University&16.43&38.5&1.07&30
\\ Swansea University&16.43&38.5&0.67&41
\\ University of Derby&11.43&40.5&1.41&12
\\ Liverpool John Moores University&11.43&40.5&0.75&40
\\ University of Glamorgan&8.57&42&0.36&43
\\ Roehampton University&7.14&43&0.36&44
\\ Bath Spa University&2.14&44&1.23&20
\\ \hline
\end{tabular}
\label{tab3}
\end{center}
\end{table}
\begin{figure}[t]
\centerline{\includegraphics[width=0.52\textwidth]{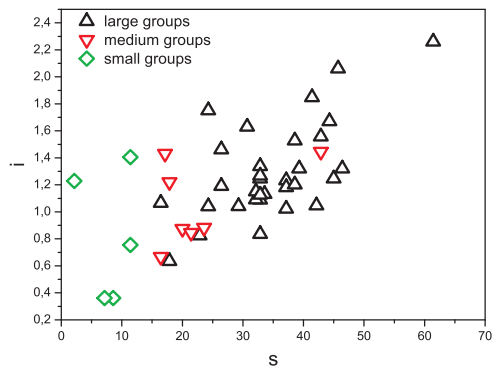}}
\caption{Correlation between $s$ (average quality of research groups according to RAE 2008) and $i$ (average excellence of research groups according to normalised citation impact).} \label{fig1_RAE_vs_Evidence}
\end{figure}
If the NCI score formed a perfect proxy for RAE peer-review
quality measures, there would be a perfect linear correlation
between  $i$ and $s$. The actual correlation is depicted in
Fig.~\ref{fig1_RAE_vs_Evidence} where $i$ is plotted against $s$
for biology research groups. The relevant data are contained in
Table~\ref{tab3}, where the RAE measured qualities and strengths
of different institutions are listed alongside
 specific and absolute NCI values. The
quality values $s$ come from RAE 2008 and the impact scores from
Ref.~\cite{Evidence_2010}. While a general alignment is evident
(groups with high  RAE  quality scores tend to have high
NCI measures) there is considerable scatter. This
is quantified by a relatively small value of the Pearson
coefficient of $64\%$. (Table~\ref{tab_coefficients} contains a
full list of Pearson correlation coefficients measured in this
paper.)

In Refs.~\cite{Ralph2010,Ralph2011}, a quantitative analysis of the dependency of group or departmental research quality on size was given.
This allowed the determination of two types of {\emph{critical mass}} in research.
It was shown that, for a multitude of different academic disciplines, there is a linear  relationship between quality and quantity up to a certain group size, known as the {\emph{upper critical mass}} $N_c$.
At this point coordination problems set in and a phenomenon similar to the Ringelmann effect\footnote{The Ringelmann effect describes the tendency for average productivity to reduce as the size of the group increases, while in Refs. \cite{Ralph2010,Ralph2011} a reduction in the ``rate of change'' of quality with quantity is observed.} ensures that groups of size greater than $N_c$ have either a reduced dependency of quality on quantity or no such dependency.
In Refs.~\cite{Ralph2010,Ralph2011} a {\emph{lower critical mass}} $N_k$ was also defined and measured for many disciplines.
This was interpreted as the minimum size a research department should achieve to be stable in the long term.
The two critical masses, the values of which are strongly dependent on the research discipline, allow research groups and departments to be categorised as being {\emph{small}} if they have size $N \leqslant N_k$, {\emph{medium}} if $N_k \leqslant N \leqslant N_c$ or {\emph{large}} if $N>N_c$.
For the biology UOA, the estimates for critical masses are $N_k = 10.4$ and $N_c = 20.8$ \cite{Ralph2010,Ralph2011}.
(Fractions of staff are a feature of RAE in that HEI's  can include part-time researchers in their submissions.
These are counted as a proportion of full time equivalence \cite{RAE_web}.)

%%%%%%%%%%%%%%%%%%%%%%%%%%%%%%%%%%%%%%%%%%%%%%%%%%%%%%%%%%%%%%%%%%%%%%%%%%%%%%%%%%%%%%%%%%%%%%%%%%%%%%%%%%%%%%%%%%%%%%%%%%%%5
\subsection{Weak correlation between specific measures of overall quality and impact}
%%%%%%%%%%%%%%%%%%%%%%%%%%%%%%%%%%%%%%%%%%%%%%%%%%%%%%%%%%%%%%%%%%%%%%%%%%%%%%%%%%%%%%%%%%%%%%%%%%%%%%%%%%%%%%%%%%%%%%%%%%%%5

The question arises whether citation counts could serve as a proxy for RAE scores for small, medium or large groups separately.
In Fig.~\ref{fig1_RAE_vs_Evidence} the small, medium and large research groups are distinguished by different colors.
While there is  considerable scatter across all three categories, it is most pronounced for small and medium groups, an observation which is  confirmed by the values of Pearson coefficients calculated separately for
small ($r<0$), medium ($r \approx 47\%$) and large groups ($r \approx 62\%$) (Table~\ref{tab_coefficients}).
Nonetheless, it provokes the question of whether size is a relevant feature which should be taken into account in attempting to build a citation-based proxy for peer review.

\begin{figure}[t]
\centerline{\includegraphics[width=0.49\textwidth]{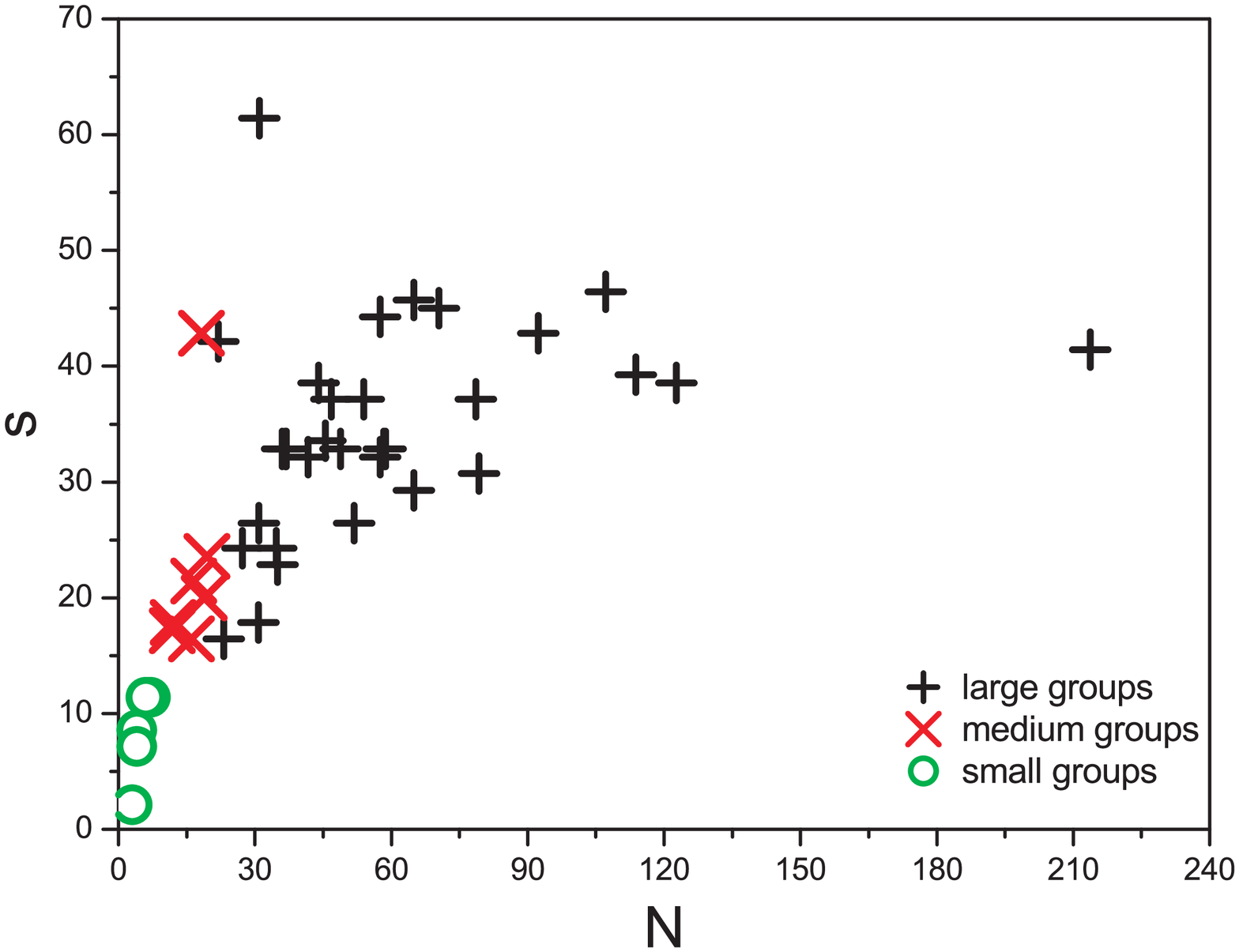}\hfill
\includegraphics[width=0.49\textwidth]{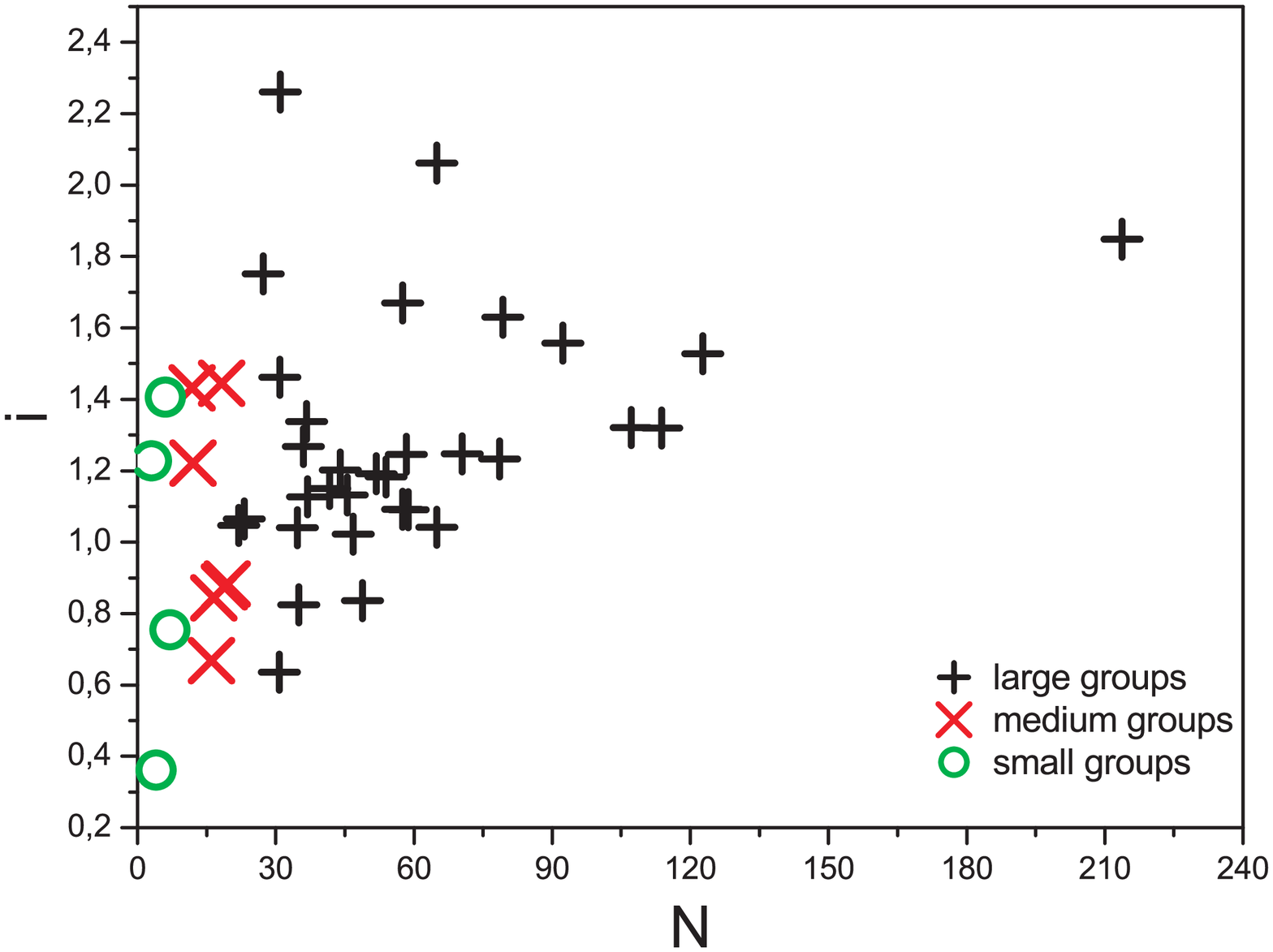}}
\centerline{(a)\hspace{8cm}(b)}
\caption{The average excellence of research groups as a function of group size $N$ for biology according to: (a) RAE 2008 assessments and (b) normalised citation impact.}
\label{fig2_quality_vs_size}
\end{figure}
The relationship between RAE-measured quality $s$ and quantity $N$
for biology is depicted in Fig.~\ref{fig2_quality_vs_size}(a) (see
also Refs.~\cite{Ralph2010,Ralph2011}). The counterpart
relationship between specific impact $i$ and quantity $N$ is shown in
Fig.~\ref{fig2_quality_vs_size}(b) (see also Ref.~\cite{Evidence_2010}). Clearly, both
quality and impact are correlated with quantity and the plots
exhibit similar features. A striking similarity is apparent
between both plots: they each have a distinct maximum and each
have no data in the bottom right regions, reflecting the fact that
there are no low quality, or low impact large biology research
groups in the UK. On the other hand, the plots differ
significantly in much of the details, especially in the region of
small and medium groups.

%%%%%%%%%%%%%%%%%%%%%%%%%%%%%%%%%%%%%%%%%%%%%%%%%%%%%%%%%%%%%%%%%%%%%%%%%%%%%%%%%%%%%%%%%%%%%%%%%%%%%%%%%%%%%%%%%%%%%%%%%%%%5
\subsection{Weak correlation between  ranked measures of overall quality and impact}
%%%%%%%%%%%%%%%%%%%%%%%%%%%%%%%%%%%%%%%%%%%%%%%%%%%%%%%%%%%%%%%%%%%%%%%%%%%%%%%%%%%%%%%%%%%%%%%%%%%%%%%%%%%%%%%%%%%%%%%%%%%%5

An increasingly common phenomenon associated with any attempt to measure scientific performance is the ranking by various media of different institutions according to their perceived quality or impact. Although overly simplistic, such tables are frequently reported  used to inform potential students of the standing of various universities. In Ref.~\cite{Ralph2010} it was shown that such systems are inherently dangerous if used to compare research quality because they do not properly take size and resources into account.

Earlier analyses  attempted to build a citation-based proxy for peer review using such rankings.
Because earlier renditions of the RAE supplied as outcome a single number (rather than a profile) purporting to encapsulate the research performance of a department,
much less information was available than we have access to in RAE 2008.
In those cases, correlations were sought not between absolute RAE scores and bibliometrics, but between {\emph{rankings}} resulting form such scores \cite{Oppenheim1996,Oppenheim2003}.
We therefore also checked for possible correlations between the ranked values of $s$ and $i$.
In order to build the ratings of research groups they should be listed in ascending order of their corresponding scores.
Each department is assigned an ascending numerical rank (the average rank in the case of the equal scores) in  Table~\ref{tab3}.
The calculated  value of the Spearman rank-order correlation coefficient is approximately equal to $57\%$ (Table~\ref{tab_coefficients})
and is even worse than the correlations based on Fig.~\ref{fig1_RAE_vs_Evidence}.

Therefore, in contrast to earlier claims in Refs.~\cite{Oppenheim1996,Oppenheim2003},
we do not observe strong correlation between rankings of HEI's according to RAE  and to  the citation-based measure.
Instead, only weak linear correlation could be observed for  ranked values of $s$ and $i$.

%%%%%%%%%%%%%%%%%%%%%%%%%%%%%%%%%%%%%%%%%%%%%%%%%%%%%%%%%%%%%%%%%%%%%%%%%%%%%%%%%%%%%%%%%%%%%%%%%%%%%%%%%%%%%%%%%%%%%%%%%%%%5
\subsection{Weak correlation between specific measures of elemental  quality and impact}
%%%%%%%%%%%%%%%%%%%%%%%%%%%%%%%%%%%%%%%%%%%%%%%%%%%%%%%%%%%%%%%%%%%%%%%%%%%%%%%%%%%%%%%%%%%%%%%%%%%%%%%%%%%%%%%%%%%%%%%%%%%%5
\label{sec23}
\begin{figure}[t]
\centerline{\includegraphics[width=0.49\textwidth]{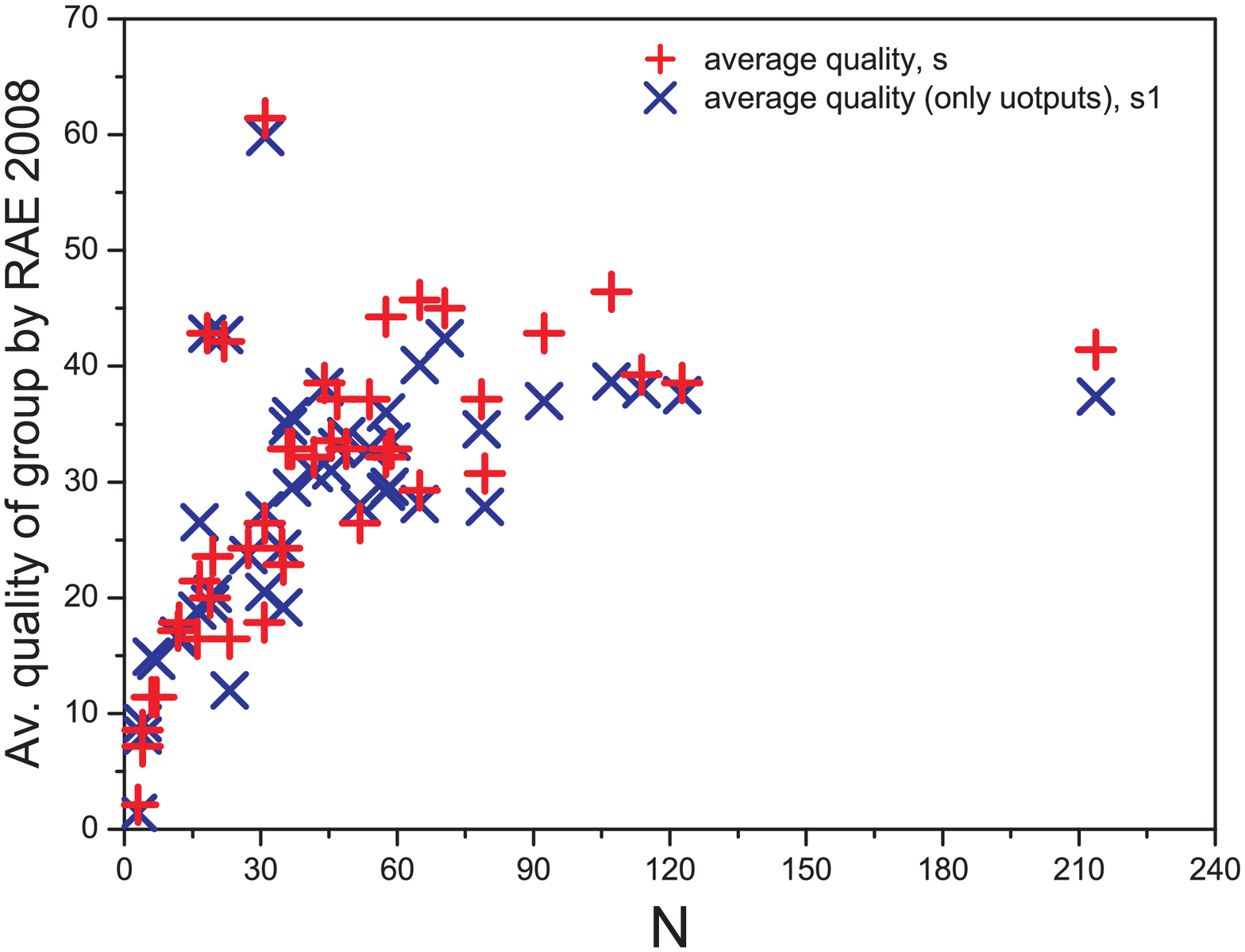}\hfill
\includegraphics[width=0.49\textwidth]{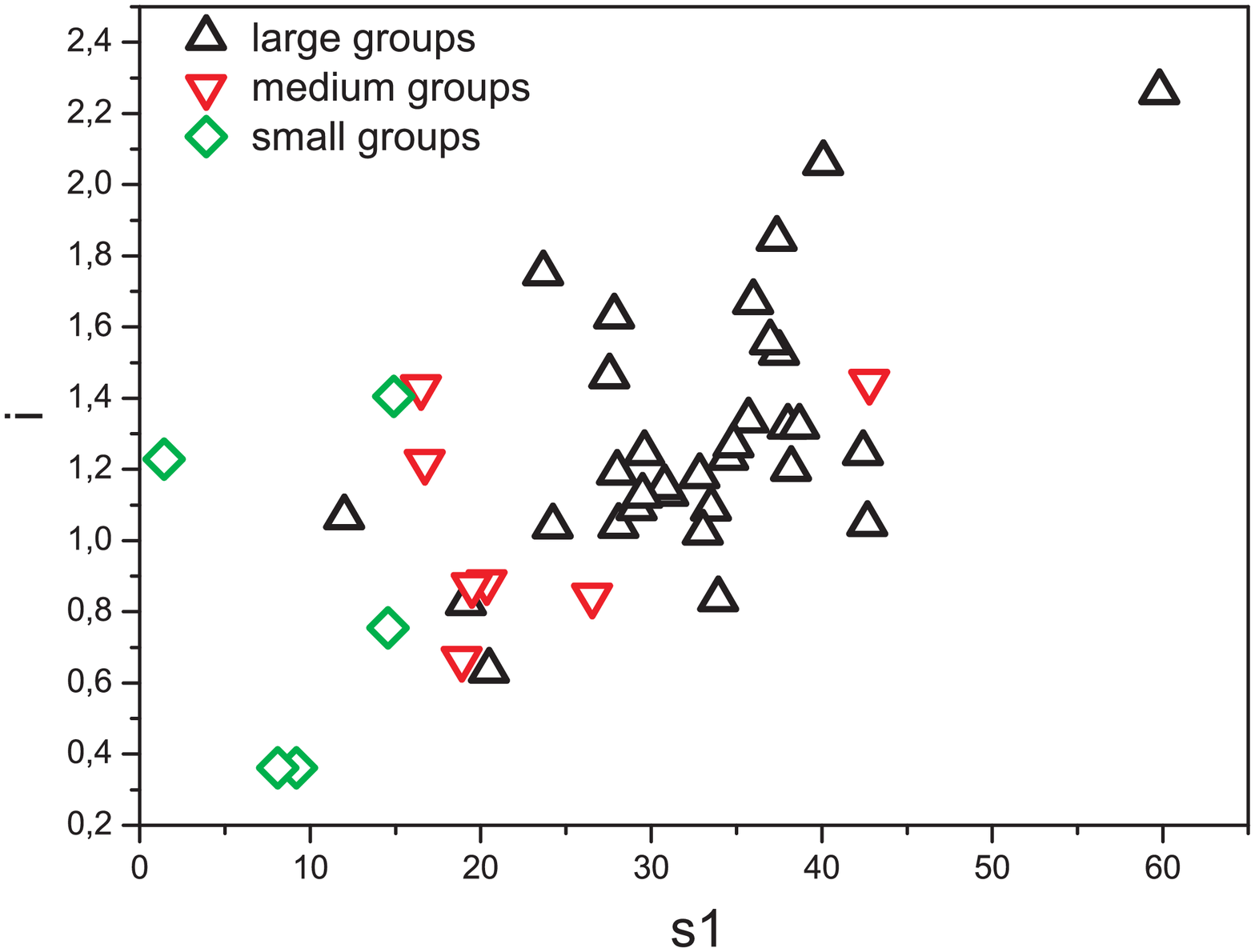}}
\centerline{(a)\hspace{8cm}(b)}
\caption{(a) The average excellence of research groups as a function of group size $N$ for biology according to RAE 2008 assessments.
Here  the overall quality $s$ corresponds to the symbols $+$, while output measures $s_1$ are denoted by $\times$.
(b) Correlation between $s_1$ (``outputs'' according to RAE 2008) and $i$ (from NCI scores).}
\label{fig3_outputs}
\end{figure}
At RAE, three elements of research quality were examined to arrive at an overall quality profile, namely {\emph{research outputs}}, {\emph{research environment}} and {\emph{research esteem}}.
For the biology UOA, to arrive at the overall quality score, these three elements were weighted with 75\% of the overall score coming from the perceived quality of outputs (mostly publications), 20\% coming from environment and 5\% attributable to esteem (see Table~\ref{tab1}).
Thus,  only 75\% of overall RAE estimations correspond to
evaluation of published papers. Citation counts, in contrast, rely
100\% on the quality of papers. Since {\emph{Evidence}} use only
the citation data to calculate NCI, we may
suppose it is more sensible to  compare these results with only
the output element or RAE. The overall quality profile, and
sub-profiles for research outputs, research environment and esteem
indicators for each submission are available on the official RAE
web-page \cite{RAE_web_file}. Here separate profile values for
``Outputs'', ``Environment'' and ``Esteem'' could be used instead
of the overall values. Hence, the separate components of the
average quality could be calculated using the same funding formula
\eqref{eq1_funding_formula}. We denote by $s_1$, $s_2$ and $s_3$
the quality measure coming from outputs, environment and esteem,
respectively.

Differences between $s$ and $s_1$ values are observable in Fig.~\ref{fig3_outputs} (a) where they are both plotted.
Obviously, however, the overall average quality scores $s$ and those coming from outputs only $s_1$ are quite close to each  other, since the contribution of $s_1$ into $s$ is 75\%.
It is interesting to observe that the overall quality score  $s$ is mainly greater than the output score  $s_1$ for large groups (see Fig.~\ref{fig3_outputs}).
This indicates that the RAE evaluators saw added benefit in large-scale research departments.
Another explanation is the extra visibility that large groups enjoy.
Small and medium groups are disadvantaged in these respects --- they have neither the visibility nor all of the facilities available to their larger competitors.

\begin{figure}[t]
\centerline{\includegraphics[width=0.49\textwidth]{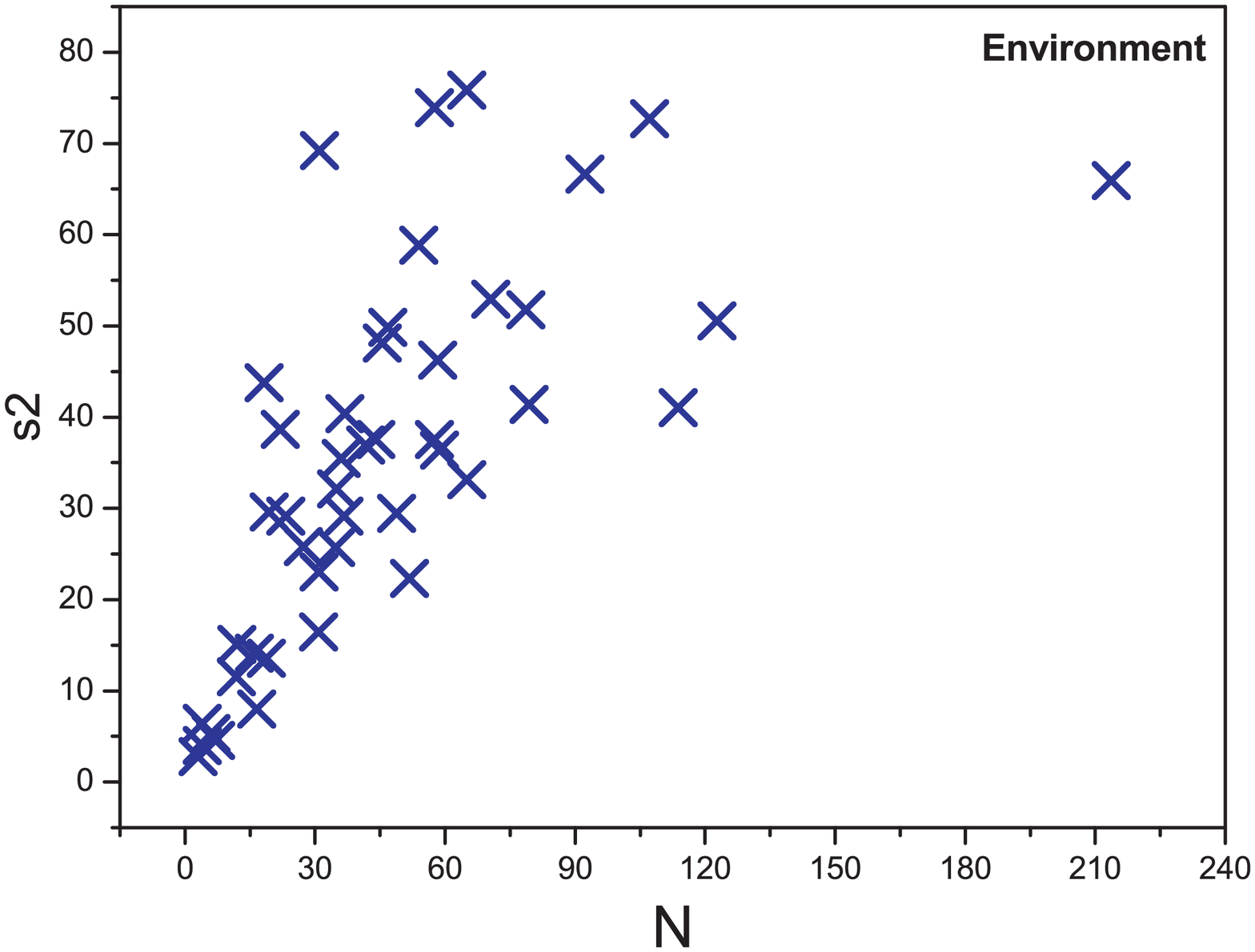}\hfill
\includegraphics[width=0.49\textwidth]{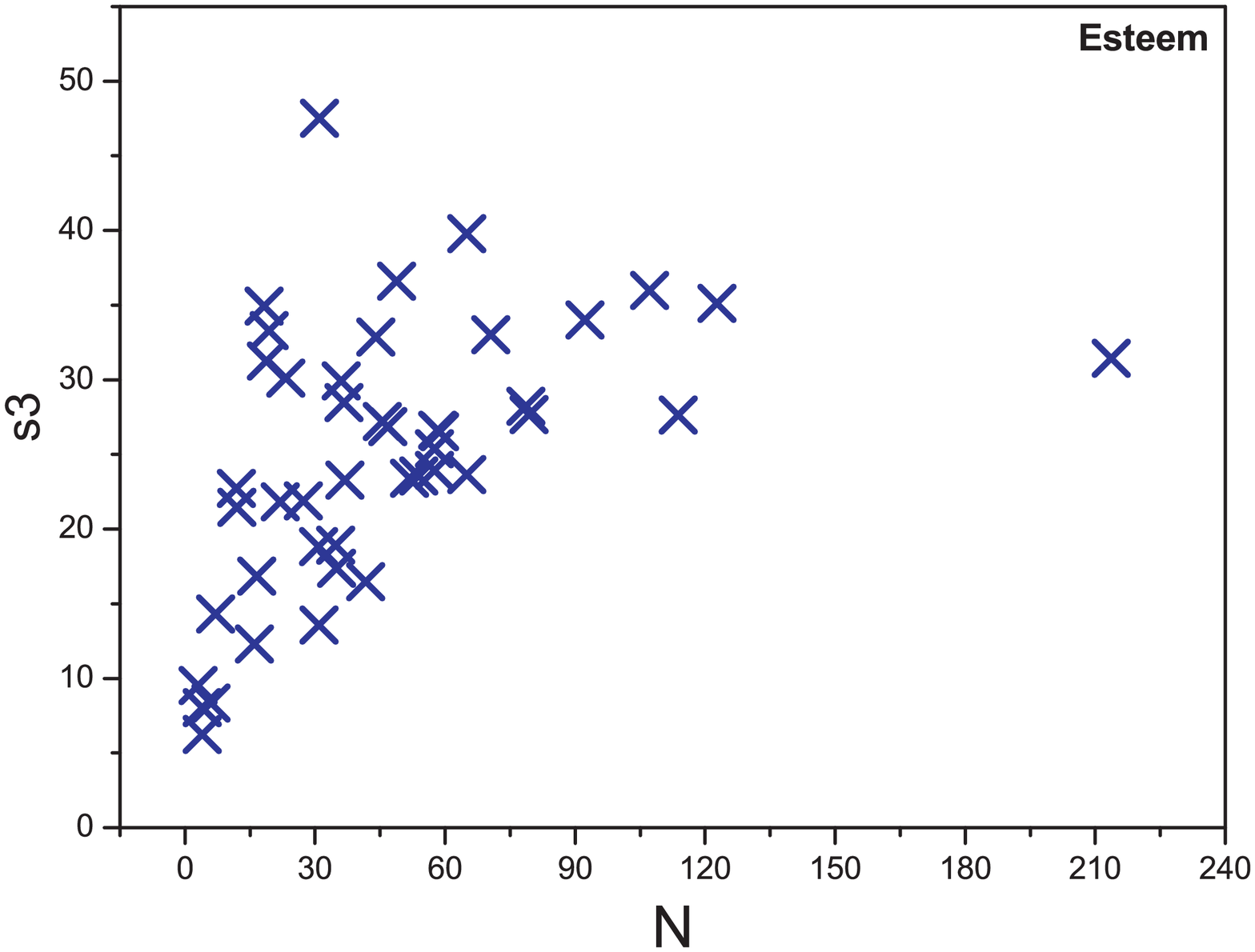}}
\centerline{(a)\hspace{8cm}(b)}
\caption{(a) Correlation between $s_2$ (``Environment'' component of average quality of research groups according to RAE 2008) and $N$. (b) Correlation between $s_3$ (``Esteem'' component of average quality of research groups according to RAE 2008) and $N$.}
\label{fig4_s2&s3_vs_N}
\end{figure}

The relationships between the ``environment'' ($s_2$) and ``esteem'' ($s_2$) components of RAE quality profiles and group size  $N$ are shown in Fig.~\ref{fig4_s2&s3_vs_N}.
One observes that $s_2$ is more linearly correlated with $N$ than the other measures $s_1$ and $s_3$.
This reflects the fact that bigger research groups have access to more expensive and more complex equipment.
Economy of scale thus ensures that the average fraction of ``environment'' continues to grow with group size, even for very large groups or departments.

Nonetheless, the linear correlation between $s_1$ and $i$ is not
stronger than that between $s$ and $i$. In fact, the corresponding
Pearson coefficient at $53\%$ is even smaller
(Table~\ref{tab_coefficients}). Moreover, there is the same high
degree of scatter for  small and several medium groups. To be
comprehensive, we have also determined the correlation
coefficients for the cases of environment quality $s_2$ and esteem
$s_3$. The resulting correlation coefficients  are $r = 55\%$ and
$r = 46\%$, respectively. In conclusion, there is only weak
correlation between {\emph{Evidence}} indicators
of impact and RAE 2008 scores for average quality of research
groups. Therefore it is not possible to use the NCI as a direct
proxy for peer review measures of academic research quality.
\begin{table}[!b]
\caption{The values of correlation coefficients calculated for different data sets.}
\begin{center}
{{
\begin{tabular}{|l|l|l|l|l|}
\hline
Description of the  & Pearson coefficient $r$ and $P$-value& Spearman \\
data sets &comparing to significance level $\alpha=0.05{}^{\dag}$& coefficient\\
\hline \hline
$s$ vs $i$& $r\approx 0.64$, $P<\alpha$&$ 0.57$\\
& (large groups) $r\approx 0.62$, $P<\alpha$&\\
& (medium groups) $r\approx 0.47$, $P>\alpha$&\\
& (small groups) $r<0$, $P>\alpha$&\\
& (small \& medium groups) $r\approx 0.39$, $P>\alpha$&\\
\hline
$s_1$ (only outputs) vs $i$&$r\approx 0.60$, $P<\alpha$&$ 0.53$\\
& (large groups) $r\approx 0.57$, $P<\alpha$&\\
& (medium groups) $r\approx 0.36$, $P>\alpha$&\\
& (small groups) $r\approx 0.03$, $P>\alpha$&\\
& (small \& medium groups) $r\approx 0.35$, $P>\alpha$&\\
\hline
$s_2$ (only environment) vs $i$&$r\approx 0.64$, $P<\alpha$&$ 0.55$\\
& (large groups) $r\approx 0.63$, $P<\alpha$&\\
& (medium groups) $r\approx 0.40$, $P>\alpha$&\\
& (small groups) $r< 0$, $P>\alpha$&\\
& (small \& medium groups) $r\approx 0.36$, $P>\alpha$&\\
\hline
$s_3$ (only esteem) vs $i$&$r\approx 0.58$, $P<\alpha$&$ 0.46$\\
& (large groups) $r\approx 0.54$, $P<\alpha$&\\
& (medium groups) $r\approx 0.40$, $P>\alpha$&\\
& (small groups) $r\approx 0.20$, $P>\alpha$&\\
& (small \& medium groups) $r\approx 0.41$, $P>\alpha$&\\
\hline
$\mathcal{S}$ vs $\mathcal{I}$&$r\approx 0.97$, $P<\alpha$&---
\\
& (large groups) $r\approx 0.96$, $P<\alpha$&\\
& (medium groups) $r\approx 0.86$, $P<\alpha$&\\
& (small groups) $r\approx 0.65$, $P>\alpha$&\\
& (small \& medium groups) $r\approx 0.92$, $P<\alpha$&\\
 \hline
\end{tabular}
}} \label{tab_coefficients}
\end{center}
\footnotesize{${}^{\dag}$\textit{if $P<\alpha$ then the linear correlation is considered as statistically significant}}
\end{table}

%%%%%%%%%%%%%%%%%%%%%%%%%%%%%%%%%%%%%%%%%%%%%%%%%%%%%%%%%%%%%%%%%%%%%%%%%%%%%%%%%%%%%%%%%%%%%%%%%%%%%%%%%%%%%%%%%%%%%%%%%%%%5
\section{Strong correlation between absolute measures of research strength and impact}
%%%%%%%%%%%%%%%%%%%%%%%%%%%%%%%%%%%%%%%%%%%%%%%%%%%%%%%%%%%%%%%%%%%%%%%%%%%%%%%%%%%%%%%%%%%%%%%%%%%%%%%%%%%%%%%%%%%%%%%%%%%%5

A conclusion of the above analysis is that there is only  weak correlation between the specific measure of impact $i$ and
the RAE peer-evaluated measure of quality $s$, and that this is especially weak, or absent for  small groups.
Possible reasons for this include the facts that (i) there are  few small and medium groups (only 5 small groups and only
only 7 medium groups) in biology and (ii) small groups, by definition, have relatively few members.
Both of these factors mean that the statistics for small (and to a lesser extent medium) groups are relatively poor in comparison to those for large groups.
Since the number of large groups is bigger and also the statistics for large groups are better, more robust results can be obtained for large groups.

Moreover, in the above analysis, all research groups are treated as contributing the same weight to the analysis.
E.g., the three researchers from Bath Spa University contribute to the same extent as the 213.69 biologists from Cambridge University!
It is therefore sensible to correct this anomaly by introducing weights.
Multiplying the average  quality of groups by their size renders the specific measures absolute: quality becomes strength and the NCI is also scaled up to the volume of the group or department.
\begin{figure}[h]
\centerline{\includegraphics[width=0.52\textwidth]{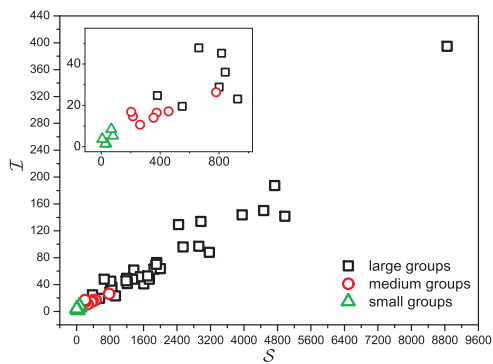}}
\caption{Correlation between absolute quality values (strength) for research groups according to RAE 2008 and normalised citation impact results (the region of small and medium groups on the inset).}
\label{fig5_RAE_vS_Evidence_unnormalized}
\end{figure}

The RAE, peer-reviewed  measures of strength ${\mathcal{S}}$ are compared to the absolute, citation-based measures of NCI in Fig.~\ref{fig5_RAE_vS_Evidence_unnormalized}.
The correlation between the two measures is impressive.
This is reflected in the almost perfect Person correlation coefficient $r = 97\%$ (see Table~\ref{tab_coefficients}).
Moreover, restricting the analysis only to large groups also reveals an excellent correlation of $96\%$.
The correlation coefficients for medium groups alone is less good, at $86\%$ and for small groups that figure is $65\%$.
However, since small and medium research groups tend to have the same linear dependency of quality on quantity \cite{Ralph2010} it is more sensible to combine them in the correlation analysis. Indeed, the Pearson correlation coefficient for small and medium groups combined is $92\%$.
Thus we can say that the NCI forms a good proxy for peer-review estimates of group research strength for large and small/medium research departments.
Fig.~\ref{fig5_RAE_vS_Evidence_unnormalized} should be compared to Fig.~\ref{fig1_RAE_vs_Evidence}.
The replacement of specific measures of quality and impact by their absolute counterparts has had the effect of stretching  the corresponding axes by an amount proportional to the quantity of the groups or departments. Because of the clear relationship between quality and quantity identified in Ref.~\cite{Ralph2010,Ralph2011}, and a similar relationship between NCI impact and quantity observed in
Ref.~\cite{Evidence_2010}, this stretching induces the improved correlations observed.

%%%%%%%%%%%%%%%%%%%%%%%%%%%%%%%%%%%%%%%%%%%%%%%%%%%%%%%%%%%%%%%%%%%%%%%%%%%%%%%%%%%%%%%%%%%%%%%%%%%%%%%%%%%%%%%%%%%%%%%%%%%%5
\section{Conclusions}
%%%%%%%%%%%%%%%%%%%%%%%%%%%%%%%%%%%%%%%%%%%%%%%%%%%%%%%%%%%%%%%%%%%%%%%%%%%%%%%%%%%%%%%%%%%%%%%%%%%%%%%%%%%%%%%%%%%%%%%%%%%%5

A goal of scientometrics is to develop a method to provide a reliable measurement of scientific excellence using minimal efforts.
However, any attempts to replace current peer-evaluation based systems have to be done carefully and robustly.
Quality related funding, such as that administered by HEFCE post RAE in the UK is proportional to absolute measures of research strength since financial support for research obviously needs to be  weighted according to group size.
Therefore, a citation based approach, at least for large groups,  may be introduced as less intrusive and more cost effective alternative to national peer review.
This would also have the advantage of ameliorating the distortions to the research system introduced on a national basis every five years or so through Goodhart's phenomenon, although one might expect a citation-based system to introduce distortions of its own.

However, we have also seen that the citation metrics used here are not well correlated with peer-review measures of group research excellence.
Since the latter are used to rank institutions, it is clear that citation counts are not a good basis on which to make such comparisons.

Thus we arrive at the almost paradoxical conclusion that the citation-based metric (NCI) may be used as an excellent proxy for peer-reviewed measurements of institutional scientific strength but it is only a poor proxy for measures of quality.
Since quality related funding is strength based, this may offer a much cheaper alternative to the system currently in use in the UK and some other countries.

The  analysis presented here is based on biology research groups in the UK. Further analyses are underway for other disciplines to determine the broader suitability of  scientometric measures of specific and absolute research-group properties.
The nuances connected with peculiarities of groups of different sizes should be also studied.
Since in the UK system, a large group with, say 40 members, provides  $40 \times 4=160$ submissions, which may be sufficiently large for statistical fluctuations to be ironed out.
But the RAE evaluation of a small group, say with 5 staff members, is based on the statistics of only approximately $5\times 4=20$ submissions.
This is far more susceptible to inaccuracies and statistical noise.
So, any automated evaluation of large groups is much more robust than for small groups -- even if one is only interested in measuring strength rather than quality.
A possible way to balance this, if introducing citation counts, would be to keep peer review for small and medium sized groups or to require a greater number of outputs per person for such groups to improve statistics.
Investigations into such schemes are underway.
In any case, it is clear from the current test case that any attempt to automatically evaluate research quality should very carefully take account of group or departmental sizes \cite{Ralph2010}.

%\newpage
%%%%%%%%%%%%%%%%%%%%%%%%%%%%%%%%%%%%%%%%%%%%%%%%%%%%%%%%%%%%%%%%%%%%%%%%%%%%%%%%%%%%%%%%%%%%%%%%%%%%%%%%%%%%%%%%%%%%%%%%%%%%5
\section*{Acknowledgements}
%%%%%%%%%%%%%%%%%%%%%%%%%%%%%%%%%%%%%%%%%%%%%%%%%%%%%%%%%%%%%%%%%%%%%%%%%%%%%%%%%%%%%%%%%%%%%%%%%%%%%%%%%%%%%%%%%%%%%%%%%%%%5
This work was supported in part by the 7th FP, IRSES project No. 269139 ``Dynamics and cooperative phenomena in complex physical and biological
environments'' and IRSES project No. 295302 ``Statistical physics in diverse realizations''.
The authors thank Jonathan Adams from Evidence for the data and Ihor Mryglod for fruitful discussions.

%%%%%%%%%%%%%%%%%%%%%%%%%%%%%%%%%%%%%%%%%%%%%%%%%%%%%%%%%%%%%%%%%%%%%%%%%%%%%%%%%%%%%%%%%%%%%%%%%%%%%%%%%%%%%%%%%%%%%%%%%%%%5

\end{document}